\documentclass[fleqn]{article}%
\usepackage{amsmath}
\usepackage{amsfonts}%
\usepackage{mathtools}
\usepackage{amssymb}%
\usepackage{graphicx}
\usepackage{stmaryrd}
\usepackage[makeroom]{cancel}
\usepackage{caption}
\usepackage{framed}
\usepackage{fancyvrb}
\usepackage{multirow}
\usepackage{listings}
\usepackage[utf8]{inputenc}
\usepackage{newunicodechar}
\usepackage{CJK}
\usepackage[a4paper, left=25mm, right=25mm]{geometry}
\usepackage{epigraph}
\usepackage{multicol}
\usepackage{semantic}
\usepackage{hyperref}

\makeatletter
\newcommand{\dotDelta}{{\vphantom{\Delta}\mathpalette\d@tD@lta\relax}}
\newcommand{\d@tD@lta}[2]{%
  \ooalign{\hidewidth$\m@th#1\mkern-1mu\cdot$\hidewidth\cr$\m@th#1\Delta$\cr}%
}
\makeatother

\VerbatimFootnotes

\begin{document}

\title{Multidimensional Predicates for Prolog}
\author{G\"unter Khyo}
\date{\today}
\clearpage\maketitle
\thispagestyle{empty}

\abstract{
In 2014, Ungar et al. proposed \emph{Korz}, a new computational model for structuring adaptive (object-oriented) systems \cite{korz}. Korz combines implicit parameters and multiple dispatch to structure the behavior of objects in a multidimensional space. Korz is a simple yet expressive model which does not require
special programming techniques such as the Visitor or Strategy pattern to accommodate a system for emerging contextual requirements.

We show how the ideas of Korz can be integrated in a Prolog system by extending its syntax and semantics with simple meta-programming techniques. We developed a library, called \emph{mdp} (multidimensional predicates) which can be used to experiment with multidimensional Prolog systems. We highlight its benefits with numerous scenarios such as printing debugging information, memoization, object-oriented programming and adaptive GUIs. 

In particular, we point out that we can structure and extend Prolog programs with additional concerns in a clear and concise manner. We also demonstrate how Prolog's unique meta-programming capabilities allow for quick experimentation with syntactical and semantical enhancement of the new, multidimensional model. 

While there are many open concerns, such as efficiency and comprehensibility in the case of larger systems, we will see that we can use the leverage of mdp and Prolog to explore new horizons in the design of adaptive systems.
}

\newpage

\section{Introduction}

% Two things missing: [foo: bar]  ? ... extends implicit context

Consider the problem of finding a path in a directional graph. The graph is represented by the \emph{edge(N1, N2)}-relation. In anticipation of changing requirements, we decide to use \emph{mdp} from the beginning and define the transitive relation \emph{path(From, To)} as follows:

\begin{lstlisting}
[] # edge(a, b). [] # edge(b, c). % ...

[] # path(A, B) :- ? edge(A, B).

[] # path(A, C) :-
  ? edge(A, B),
  ? path(B, C).
\end{lstlisting}

As a multidimensional relation, \emph{path/2} consists of two parts separated by a hashtag: (1) a list of concerns (its context) and (2) the name of the predicate. Currently, there are no concerns and therefore all context specifications are empty. For implementation reasons, when evaluating a multidimensional relation the \verb|?| operator has to be supplied explicitly.

Suppose we want to add debugging output to our program. We add a new variant of \emph{edge/2} by adding a debugging dimension to its context:

\begin{lstlisting}
[debug: P] # edge(A, B) :-
   [-debug] ? edge(A, B), % Remove debug dimension from implicit context
   apply(P, [(A, B)]).
\end{lstlisting}

We get the desired behavior by executing \emph{path/2} within a debug context,
i.e., 

\verb|[debug: writeln] ? path(X, Y)|. The context is implicitly passed along the 
call chain. If there are multiple definitions of a predicate, the definitions which have the highest number of matching dimension/value pairs will be chosen. Thus, our augmented definition of edge/2 will be preferred over the plain definition.

In addition, the programmer may add arbitrary Prolog expressions to a context specification. These then serve as preconditions which have to hold for the matching result to count. For example, we could constrain our debugging context as follows:

\begin{lstlisting}

[debug: P, is_io_predicate(P, IODevice), ready(IODevice)] # edge(A, B) :- ...

% definitions for is_io_predicate/2 and ready/1 ...
\end{lstlisting}

(where the programmer has to provide definitions for \emph{is\_io\_predicate} and \emph{ready} which serve as illustration purposes only in this example).

If any of the preconditions do not hold, the default variant with no debugging will be executed. If we do not want this behavior but throw an exception instead, we have to move the preconditions inside the body of \emph{edge} and throw an exception.

While we have a clean solution for debugging, we have to write our debugging code for every predicate, which will soon become tedious. We can generalize our debugging concern for arbitrary predicates by defining an anonymous multidimensional rule:

\begin{lstlisting}
[ predicate: P, 
  debug: D, 
  is_io_predicate(P, IODevice), ready(IODevice)] :-
	
	[-debug] ? P,
	apply(D, [P]).
\end{lstlisting}

The dimension \emph{predicate} is part of every context and is supplied by mdp automatically. Note that it is important that we remove our specialized debugging rule for \emph{edge/2}, otherwise, we execute the generalized definition and the specialized rules since both have equal matching scores \footnote{The programmer may also influence this behavior by assigning weights to context specifications (an example is shown in Section~\ref{sec:subtyping}).}.

Looking back at the problem of finding paths, as experienced programmers we know immediately that our solution will only work for acyclic graphs. To accommodate for cycles, we have to keep record of already visited nodes. However, instead of simply adapting our implementation to the most general case, we decide to keep the simpler and more efficient implementation for acyclic graphs. We make following arrangements:

\begin{lstlisting}
[] # path(A, B) :- ? edge(A, B).

[graph_type: acyclic] # path(A, C) :-
  ? edge(A, B),
  ? path(B, C).
\end{lstlisting}

We then add following lines:

\begin{lstlisting}
[graph_type: cylic, nodes: _] # path(A, B) :- ? edge(A, B).

[graph_type: cylic, nodes: L] # path(A, C) :-
  ? edge(A, B),
  \+ member(B, L),

  append([A], L, LWithA),
  [nodes: LWithA] ? path(B, C).  

[graph_type: cyclic] # path(A, C) :-
  ? edge(A, B),
  [nodes: []] ? path(B, C).
\end{lstlisting}

When we state the query 

\begin{lstlisting}
?- graph_type(T), [graph_type: T] ? path(X, Y)
\end{lstlisting}

the most efficient of the available implementations will be chosen. % wording

However, we have to be careful: If we do not provide any context information to our query, our program may yield unpredictable results or even hang (since all path definitions are equally specific and get "`mixed-up"'). For the same reason, it is also crucial that the first line also includes the \verb|nodes| dimension even if it is not needed in the base-case. 

To fix this, in the default case, if we know nothing about the structure of a graph, it is safe to execute the predicate which can handle cycles. We add following line:

\begin{lstlisting}
[] # path(A, B) :- [graph_type: cyclic] ? path(A, B).
\end{lstlisting}

There are many other aspects that we can improve on. For example, we can add another dimension which describes the type of data-structure for holding the visited nodes. Instead of using a list, we could use a set or any data structure and let mdp select the most efficient predicates for inserting and testing for members. Expanding the example is straightforward and left as an exercise to the reader.

In summary, \emph{mdp} allows us to add concerns to a prolog program with relative ease. There are many other useful applications for mdp as shown in Section~\ref{sec:examples}, where we will see how we can model objects, context-sensitive GUIs and memoization. Details on the implementation and guidance on how to extend mdp are found in Section~\ref{sec:implementation}.

\newpage

\section{Further mdp-Examples}
\label{sec:examples}

In this section we will see more examples that motivate the use of mdp. We will address following issues: memoization, object-oriented programming and how we can augment a GUI with additional concerns.

\subsection{Memoization}

Memoization is a common technique used to cache expensive computations. In this section, we will explore how we can orthogonally add the concern of memoization to prime number generation. We will then derive a generic solution to memoize arbitrary predicates.

Suppose we have a naive implementation of a prime number test:

\begin{lstlisting}
[] # is_prime(P) :-
   P > 1,
   UpperTestLimit is floor(sqrt(P)),
   forall(between(2, UpperTestLimit, Divisor), \+ 0 is P mod Divisor).  
\end{lstlisting}

Fortunately we know that, for some reason, there are only a few unknown numbers we are testing so we might get away with memoization. We add following lines:

\begin{lstlisting}
:- dynamic memoized_is_prime/2.

[memoize: _] # is_prime(P) :-
  memoized_is_prime(P, IsPrime) ->
    call(IsPrime) ;
   (
       [-memoize] ? is_prime(P) ->
          assertz(memoized_is_prime(P, true)) ;
          assertz(memoized_is_prime(P, false))
   ).		
\end{lstlisting}

This works fine, but if we have to copy and paste the memoization code for every other predicate we would like to enhance with memoization. We remove our specialized implementation and generalize as follows: 

\begin{lstlisting}
:- dynamic memoized/2.

[memoize: _, predicate: P, memoized(P, TorF)]  :- call(TorF).

[memoize: _, predicate: P, \+ memoized(P, _)] :-
	forall([-memoize] ? P, assertz(memoized(P, true))),

	(
		\+ [-memoize] ? P -> 
			assertz(memoized(P, false)), fail ;
			memoized(P, _)
	).
\end{lstlisting}

This solution is slightly more complex, since it also handles memoization for non-deterministic predicates. Now, for every multidimensional predicate $P$ we can state the query:

\begin{lstlisting}
?- [memoize: _] ? P(Args)
\end{lstlisting}

to memoize. Our solution can be further improved by extending \emph{assert} and \emph{retract} to invalidate memoization data, which is left as an exercise to the reader.

\subsection{Object-oriented Programming with mdp}
\label{sec:object-oriented-programming}

In this section, we will explore how mdp can be used to implement a purely object-oriented system, i.e., no classes. To this end, we built a simple meta-object-protocol \emph{mop} on top of mdp.  
In the following sections, we explore how we can represent messages and state in mdp and implement a model for subtyping, polymorphism and protecting data.

\subsubsection{Objects, Messages and State} 

In mop, as in Korz, an object is a globally unique identifier. New objects (identifiers) are created with the predicate \verb|new_oid(+NewOID)|. Messages can be sent to objects with the \emph{!} operator (inspired by Erlang), for example, \verb|Display ! render(Object)| which is equivalent to \verb|[rcvr: Display] ? render(Object)|.  Thus, a method is just a multidimensional predicate which includes a receiver dimension, containing the id of the object. 

The attribute of an object can be inspected and changed with \verb|OID ! read(Name, Value)| and \verb|OID ! write(Name, NewValue)|, respectively. Also, for every object there are predicates to clone the object (which is a shallow copy of its attributes) and to query its type (which, internally, is just an attribute).

The meta-object protocol is defined as follows:

\begin{lstlisting}
:- dynamic data/3. % data(OID, Name, Value)

:- op(0101, xfx, !). % Send a message

[rcvr: OID] # write(Name, Value) :-
  retractall(data(OID, Name, _)),
  assertz(data(OID, Name, Value)).
  
[rcvr: OID] # read(Name, Value)  :- data(OID, Name, Value).

[rcvr: OID] # type(T) :- data(OID, type, T).

[rcvr: OID] # clone(OIDClone) :-
  new_oid(OIDClone),
  forall(data(OID, Name, Value), OIDClone ! write(Name, Value)).
\end{lstlisting}

The message-send operator is implemented using a hook provided by \emph{mdp} (see
Section~\ref{sec:implementation}):

\begin{lstlisting}
% Translate Receiver ! P(Args) to [rcvr: Receiver] ? P(Args)
hook_mdp_term(
  ImplicitContext,
  Receiver ! Predicate,
  ?(ImplicitContext, [rcvr:Receiver], Predicate)).

% Translate [Context] ? Receiver ! P(Args) to 
%           [rcvr: Receiver, Context] ? Receiver ! P(Args)
hook_mdp_term(
  ImplicitContext,
  Receiver ! (?(GivenContext, Predicate)),
  ?(ImplicitContext, [rcvr: Receiver | GivenContext], Predicate)).
\end{lstlisting}

The last definition allows us to provide a context to a message-send (we will see examples of this in Section~\ref{sec:protecting-data} and Section~\ref{sec:adaptive-gui}). (There are other hook-definitions to implement further syntactic variations which can be found in the mdp distribution.)

\subsubsection{Subtyping and Polymorphism}
\label{sec:subtyping}

Let us consider the canonical example of modeling objects and behavior for representing geometric shapes. On the type level, we define following subtype-relations:

\begin{lstlisting}
subtype(shape, rectangle).
subtype(rectangle, special_rectangle).
subtype(shape, circle).
\end{lstlisting}

We define the polymorphic predicate (method) \emph{representation/1} as follows:

\begin{lstlisting}

% OID < shape reads: OID is a subtype of shape
[rcvr: Shape, Shape < shape] # representation(_) :-
	throw('Abstract method: No implementation').

[rcvr: Rectangle, Rectangle < rectangle] # representation(R) :-
	Rectangle ! read(width, W),
	Rectangle ! read(height, H),
	R = rectangle(W, H).
	
[rcvr: Rectangle, Rectangle < special_rectangle] # representation(R) :-
	Rectangle ! read(width, W),
	Rectangle ! read(height, H),
	R = special_rectangle(W, H).
	
[rcvr: Circle, Circle < circle] # representation(R) :-
	Circle ! read(radius, Radius),
	R = circle(Radius).

\end{lstlisting}

Internally, the subtype-relation \verb|<| is translated to a form mdp can understand, i.e.,

\begin{lstlisting}
[rcvr: OID, OID < rectangle] # representation(R) :- ...

% Translates to

[rcvr: OID, 
  [rcvr: OID] ? type(ActualType), 
  @(type_affinity(Subtype, ObjectType, D), D)
] # representation(R) :- ...
\end{lstlisting}

We can specify this translation by defining a rule for a hook supplied by mdp:

\begin{lstlisting}
hook_context_rule_mdp_term(Context, OID < Subtype, Translation) :-
  Translation = (
    [rcvr: OID] ? type(ObjectType),
    @(type_affinity(Subtype, ObjectType, D), D)
  ).
\end{lstlisting}

where the predicate \verb|type_affinity| measures how close the supplied type constraint fits the actual type of the objects. The result is then used to give weight to the rule, using the $@$-annotation, to ensure proper matching.

\begin{lstlisting}
type_affinity(T, S, N) :-
	max_type_distance(D),
	type_distance(T, S, DistanceTS),
	N is D - DistanceTS + 1.

type_distance(T, T, 1).

type_distance(T, S, N) :-
	T \= S,
	subtype(Parent, S),
	type_distance(T, Parent, N1),
	N is N1 + 1.

max_type_distance(D) :-
	findall(D, (
              subtype(T, _), 
              subtype(_, S), 
              type_distance(T, S, D)), Distances),
	max_list(Distances, D).

\end{lstlisting}

While this is a highly inefficient implementation for a subtype check, we can use memoization to make the type check efficient and we have the advantage of a readable specification.

We can now run the query \verb|AnyShapeType ! representation(R)| and we get the representation corresponding to the type of $AnyShapeType$. Since there are no classes, prototypes have to be defined and cloned to create new instances. For example, the rectangle prototype can be created as follows: 

\begin{lstlisting}
?-	new_oid(Rectangle), 
	Rectangle ! write(type, rectangle),
	Rectangle ! write(width, 100),
	Rectangle ! write(height, 100),
	
	Rectangle ! representation(R).
	
R = rectangle(100, 100).
\end{lstlisting}

There are many improvements we can make to our meta-object protocol. For example, we could check whether an attribute has been declared before writing to it. 

\subsubsection{Protecting Data}
\label{sec:protecting-data}

In many class-based object-oriented languages, the programmer may restrict access to certain attributes or methods with visibility modifiers. For example, consider following game:

\begin{lstlisting}
% Overrides default mop read method
[rcvr: OID, OID < guessing_game] # read(secret_number, S) :-
  throw('No funny stuff').

[rcvr: OID, OID < guessing_game, modifier: private] # read(secret_number, S) :-
  number(S),
  data(OID, secret_number, S).

[rcvr: OID, OID < guessing_game] # guess(S) :-
  [modifier: private] ? OID ! read(secret_number, S).
	
\end{lstlisting}

We could improve our solution by protecting access to \emph{data/3} and providing a cryptographic context which cannot be tempered with. The reader is encouraged to experiment with more secure models.

The key point is that mdp provides all facilities to model a variety of concerns and no special language constructs such as visibility modifiers are needed. In fact, mdp offers even more flexibility, for instance, we can also imagine role-based visibility.

\subsubsection{But why are there no classes?}

Most object-oriented languages rely on concepts such as classes, meta-classes, subtyping and visibility modifiers for data protection. In accordance to the research of Self \cite{self, object-focus}, we do not believe that there should be a separate language concept for classes (or a fixed class model) for following reasons: 

\begin{enumerate}

\item Classes as abstract entities break with the real-world metaphor of having tangible objects. Humans are better dealing with the concrete, rather than abstract, non-tangible ideas.

\item Classes introduce the problem of infinite meta-regress \cite{self}. If there is a class, there must be a meta-class, and if there is a meta-class, there must be a meta-meta-class and so on.

\item A purely object-based system can be used to implement a class-based system. For example, a Smalltalk interpreter can be written in Self.

\item Just as classes, objects can serve as namespaces, protect access to attributes and do everything a class description can do, even more.

\end{enumerate}

mdp gives us the flexibility to create objects and contexts that serve as class descriptions or in general, ontologies of any kind. The reader is encouraged to experiment with new ideas using mdp and mop.

\subsection{An Adaptive GUI}
\label{sec:adaptive-gui}

In this section we will explore how we can model an adaptive graphical
user interface system. (Note that the example is only a sketch and serves for demonstrating the potential of mdp. We will provide full-fledged examples of adaptive GUIs in future works).

The system executes within a dynamic context which can be manipulated by the user or other parts of the system. For the moment, we do not make any assumptions how the context is structured, i.e., which dimensions of user-experience it contains. In this example, we focus on displaying objects only.

In some part of the system, assume we have a predicate \emph{refresh(Display, Object)} which is evaluated whenever there is a change in the system context or one of its graphical objects:

\begin{lstlisting}

refresh(Display, Object) :-
	system_context(Context),
	Context ? Object ! representation(R),
	Display ! present(R).

\end{lstlisting}

Different types of objects have different graphical representations. For the moment, we assume that all objects are represented as colored bitmaps. We use a third-party library for rasterization. There might be following implementations for the predicate \emph{representation}.

\begin{lstlisting}

[rcvr: G, G < text] # representation(R) :-
	G ! text_contents(T),
	G ! text_color(Color),

	text_pixmap(T, Color, R). % Third party library
	
[rcvr: G, G < box]  # representation(R) :-
	G ! bounds(Width, Height),
	G ! color(Color),
	
	box_pixmap(Width, Height, Color, R). % Third party library
	
% Specifications for other shape types ...

% If we do not recognize a new type of object, use a default representation
[rcvr: G, G < object] # representation(R) :-
	default_representation(R).

\end{lstlisting}

%Note that the predicate which is most specific to its context will be selected. The least %specific predicate, defined in the last line, will be selected whenever the other contexts %do not match.

Many modern GUIs such as browsers accept different types of graphical representations, for example vector graphics. Assuming we find out that our display object is capable of rendering vector graphics, we do not have to change any of the existing code to support this. We just add the dimension \emph{render\_type} to our system (and provide means of render-type-selection for the user) and add a corresponding predicate definition: 

\begin{lstlisting}

[rcvr: G, G < box, render_type: svg] # representation(R) :-
	G ! bounds(Width, Height),
	G ! color(Color),
	R = svg(shape=box, color=Color).
	
[rcvr: G, G < box, render_type: pixmap] # representation(R) :-
	% Fallback to original implementation
	[-render_type] ? G ! representation(R).

\end{lstlisting}

Now suppose that there is a light-sensor hooked to our system and we would like to take advantage of this by adapting the colors of the objects to the ambient light. For example, when the room is dark, we want to display our objects in a midnight blue theme. Now, we could add following lines:

\begin{lstlisting}

[rcvr: G, G < text, ambient_light: dark] # representation(R) :-
	G ! set_color(gray),
	[-ambient_light] ? G ! representation(R).
	
[rcvr: G, G < box, ambient_light: dark] # representation(R) :-
	G ! set_color(midnight_blue),
	[-ambient_light] ? G ! representation(R).	

\end{lstlisting} 

Unfortunately, this will introduce non-determinism if the context contains both an \emph{ambient\_light} and a \emph{render\_type} dimension.
The reason is that, given a graphical object, the light-sensitive and the render-type sensitive implementations of \emph{representation} match the context as none
is more specific than the other. \emph{mdp} will evaluate both variants, in the order of their definitions.

\begin{lstlisting}
?- box_prototype(T), [ambient_light: dark, render_type: svg] ? T ! representation(R).

	R = svg(shape=box, color=midnight_blue) ;
	R = svg(shape=box, color=original_color).

\end{lstlisting}

We can fix this in two ways: We can either add the \verb|render_type| dimension to our definitions or we can assign a higher score to the context: 

\begin{lstlisting}
[rcvr: G, G < box, ambient_light: dark, render_type:_] # representation(R) :-
	G ! set_color(midnight_blue),
	[-ambient_light] ? G ! representation(R).	
	


% Alternative implementation

dimension_weight(ambient_light, 2).

[rcvr: G, G < box, 
 ambient_light: dark, 
 dimension_weight(ambient_light, W)@W] # representation(R) :-
	G ! set_color(midnight_blue),
	[-ambient_light] ? G ! representation(R).	

\end{lstlisting}

However, both solutions have the problem that all variants of representation
have to be examined to safely add new variants. Nonetheless, the latter solution of assigning weights to dimension, gives us more flexibility as it allows us to organize dimension as a separate concern. Nonetheless, the programmer might forget that there is a weight assignment for a specific dimension and manually adding a weight annotation to each dimension is too tedious. However, it is straightforward extend mdp with automatic annotations.

The bigger issue is how the programmer can be assisted in structuring a multidimensional predicate space. Also, Ungar et al. stress the importance of having equal dimensions ("`no dimension holds sway over another"'). Future research and experimentation with larger systems is needed to assess this problem.

%\subsection{Controlling Side-Effects (TODO)}
%
%... Intro ...
%
%\begin{lstlisting}
%[] # assertz(P)    :- error('No permission to modify database').
%[] # retract(P)    :- error('No permission to modify database').
%
%[permission: modify-db] # assertz(P) :- assertz(P).
%[permission: modify-db] # retract(P) :- assertz(P).
%
%[] # open :- error('No permission to access file system')
%
%[permission: file-system-access] # open(SrcDest, Mode, Stream, Options) :-
  %open(SrcDest, Mode, Stream, Options).
%\end{lstlisting}
%
%Unfortunately, this solution is not waterproof since the programmer may still call a predicate the conventional way (in future works, we plan to make every predicate multidimensional by adopting a Prolog VM). Also, the permission context can be changed anywhere. A secure solution would require the use of a cryptographic context which cannot be tempered with.
%Nonetheless, the scheme might prove useful for structuring effectful prolog programs in a more comprehensible manner. 

\newpage

\section{Implementation}
\label{sec:implementation}

The implementation consists of two parts: an extensible transformer which translates mdp specifications into ordinary Prolog rules and a dispatcher which is responsible for evaluating multidimensional queries by selecting and evaluating the most specific predicates w.r.t. a given a context. 

Both the transformer and the dispatcher are implemented in standard Prolog. The transformer implements the predicate \emph{term\_expansion/3} which is offered by Prolog systems for rewriting consulted programs. 

\subsection{Transforming mdp-Code}

For every mdp rule, the transformer generates two parts, an implementation for the predicate and a signature which contains various meta-data used by the dispatcher to determine which implementation to select. 

For example, consider following mdp rule:

\begin{lstlisting}

[debug: P, weight(debug, D)@D] # edge(A, B) :-
   [-debug] ? edge(A, B), % Remove debug dimension from implicit context
   call(P, (A, B)).
	
\end{lstlisting}

The transformer will generate following code:

\begin{lstlisting}
mdp_signature(
	edge/2,  
	mdp_implementation(edge/815),
	% where 815 is a generated identifier to distinguish mdp predicates
	% that have the same functor and arity
	
	context_spec(
		% Context variable
		Ctx,
		
		% Defined dimensions
		[debug],
		
		% Translated context specification (used by the dispatcher)
		[ctx_member(Ctx, debug, P), weight(debug, D)], 
		
		% Matching-score variables (summed up by the dispatcher),
		[D]   
	)
).



mdp_implementation(edge/815, Ctx, A, B) :-
	ctx_member(Ctx, debug, P),
	? (Ctx, [-debug], edge(A, B), % invoke dispatcher
	call(P, (A, B)).

\end{lstlisting}

The basic transition rules for the transformer and the semantics of mdp are shown in Table~\ref{fig:mdp-semantics}. 

The transformer is applied to every mdp definition and executes within an environment $\Gamma$. $\Gamma$ comprises four kinds of information: $\Gamma_{context}$, $\Gamma_{dimensions}$, $\Gamma_{scores}$ and $\Gamma_{rules}$. $\Gamma_{context}$ denotes a fresh variable which holds the context of an mdp rule, $\Gamma_{dimensions}$ and $\Gamma_{scores}$ are initially empty lists that are extended by the transition rules to contain the required dimension names and the weights of an mdp rule. The weights are either variables or integer constants. Finally, $\Gamma_{rules}$ contains the translated code of a context-specification. For readability reasons its derivation is omitted.

The rules can be extended to accommodate for syntactical and semantical enhancements such as we have seen for object-oriented programming (see  Section~\ref{sec:object-oriented-programming}). For this purpose, the transformer provides two hooks that allow the extension of the basic rules. 

The first hook \emph{hook\_context\_rule\_mdp\_term(+Context, +Rule, -Term)} is applied to every term in a context specification and can be extended by the programmer to add transformation rules. $Term$ can be either ordinary Prolog code or an mdp query.

The second hook \emph{hook\_mdp\_term(+Context, +Term1, -Term1)} is applied after the first hook and transforms the context specification terms and the bodies of predicates into ordinary prolog terms.

\newcommand{\impl}[0]{\mathbin{\text{:--}}}
\newcommand{\hash}[0]{\mathbin{\#}}
\newcommand{\cons}[0]{\mathbin{|}}
\newcommand{\vsp}[0]{\\ \\[2ex]}

\begin{table}[!ht]
\centering

\begin{tabular}{|c|}

\hline \\

(Phase 1) The following rules are applied to every term in an mdp rule

\vsp

\inference[multidimensional query]
	{\Gamma \vdash Context ? P(Args)}
	{\Gamma \vdash dispatch(\Gamma_{context}, Context, P(Args))}		

\vsp

\inference[term extension]
	{\Gamma \vdash T, hook\_mdp\_term(\Gamma_{context}, T, T')}
	{\Gamma \vdash T'}	
	
\vsp

(Phase 2) Translation of mdp rules to single dimensional rules

\vsp

% dimension match
\inference[dimension]
	{\Gamma \vdash [ Dim : Coord \cons R ] \hash P(Args) \impl Body}
	{\Gamma, \Gamma_{dimensions} \cup \{Dim\} \vdash R \hash P(Args) \impl ctx\_member(\Gamma_{context}, Dim, Coord), Body}
	
\vsp

\inference[precondition]
	{\Gamma \vdash [Pre(Args_p) \cons R ] \hash P(Args) \impl Body}
	{\Gamma \vdash R \hash P(Args) \impl Pre(Args_p), Body}

\vsp
	
\inference[weight]
	{\Gamma \vdash [\_@W \cons R ] \hash P(Args) \impl Body}
	{\Gamma, \Gamma_{scores} \cup \{W\} \vdash R \hash P(Args) \impl Body}	

\vsp

\begin{tabular}{cl}
\inference[spec extension]
	{\Gamma \vdash [SpecPart \cons R ] \hash P(Args) \impl Body \\
	  hook\_context\_rule\_mdp\_term(\Gamma_{context}, SpecPart, SpecPart')}
	{\Gamma  \vdash R \hash P(Args) \impl SpecPart', Body}	&
	
(might require re-application \\ & of Phase 1) 	
\end{tabular}

\vsp

\inference[rule generation]
	{\Gamma \vdash [ ] \hash P(Args) \impl Body}
	{
	
	  \begin{array}{@{}c@{}} 
		  mdp\_implementation(P/NewID, Args, Body) \\ 
			mdp\_signature(P/Arity, mdp\_implementation(P/NewID), \\ context(\Gamma_{context}, \Gamma_{dimensions}, \Gamma_{rules}, \Gamma_{scores}))
		\end{array}
	  %\text{generate mdp\_implementation and mdp\_signature}
	}
	
\vsp

\inference[fact extension]
	{\Gamma \vdash ContextSpec \hash P(Args)}
	{\Gamma \vdash ContextSpec \hash P(Args) \impl \textbf{true}}
	
\vsp

\inference[anonymous rule]
	{\Gamma \vdash ContextSpec \impl Body}
	{\Gamma \vdash ContextSpec \hash \text{'\$anonymous\_rule'}(Args) \impl \textbf{true}}
	
\vsp
		
\hline
\end{tabular}

\caption{Overview of mdp transformation rules}
\label{fig:mdp-semantics}

\end{table}

\subsection{Predicate Selection and Evaluation}

When given a multidimensional query, the dispatcher will manipulate the implicit context (such as remove or change dimensions) and select the most specific predicates (MSPs). 

The MSPs are determined by following steps:

\begin{enumerate}
\item Collect the list of all defined mdp rules that match the functor and arity of a query. 
\item Add all anonymous rules to the list.
\item Remove all rules which have failed preconditions.
\item Remove all rules which require more dimensions than are present in the implicit context.
\item Determine the score of every rule by counting one for each matched dimension and adding up all weights.
\item Select all predicates with the highest score.
\end{enumerate}

Listing~\ref{lst:implementation} shows how the dispatcher works in detail.

\begin{lstlisting}[float, frame=single, caption=Dispatcher Implementation, label=lst:implementation, captionpos=b]

% GivenContext refers to the explicit context 
% given in a multidimensional query
dispatch(ImplicitContext, GivenContext, Predicate) :-
  % Update context information (remove or update dimensions, 
  %	                           see source distribution for details)
  updated_context(
    ImplicitContext, GivenContext, Predicate, UpdatedContext),
  
  Predicate =.. [_|Arguments],
  most_specific_predicate(UpdatedContext, Predicate, MSP),
  evaluate(MSP, UpdatedContext, [UpdatedContext|Arguments]).

most_specific_predicate(Ctx, Predicate, MSP) :-
  functor(Predicate, Name, Arity),
  
  findall(ScoreKey-MSP,
        (
          (
            mdp_signature(Name/Arity, MSP, S) ;
            mdp_signature('$anonymous_rule'/_, MSP, S)
          ),
          
          predicate_score(Ctx, S, Score),

          ScoreKey is -Score
        ),
        MSPByScore),
        
   keysort(MSPByScore, [-(Score, _)|_]),
   member(Score-MSP, MSPByScore).
	
predicate_score(Context, Signature, Score) :-
  Signature = 
    context_spec(Context, RequiredDimensions, ContextRules, Vars),
  
  ctx_keys(Context, GivenDimensions),
  intersection(RequiredDimensions, GivenDimensions , RequiredDimensions),
  length(RequiredDimensions, NumMatchingDimensions),
  
  call(ContextRules), % Evaluate preconditions
  sum_list(Vars, VarScore),
  
  Score is VarScore + NumMatchingDimensions.	
\end{lstlisting}

\newpage

\section{Related Work}

Numerous methods and techniques such as Aspect-Oriented Programming (AOP), Subject-oriented programming \cite{subject-oriented-programming, subject-topia} and special programming techniques for one-dimensional systems in the form of architectural and design patterns
have been conceived to handle concerns in adaptive systems.
The problems with these techniques are that they either require the use of special tools and languages, such as aspect description languages and compilers, or reprogramming of the mechanisms mdp already provides for every problem instance. We further argue that patterns in particular reduce the readability of context-sensitive systems since they offer no real separation between specification and implementation of concerns at the language level.

Korz, on the other hand, has been designed with multidimensionality from the start and firmly builds on the principles of uniformity and minimality. As such, the system is simple yet very expressive and readable. However, it has to be pointed out that techniques such as AOP, are in general, more expressive (but not necessarily as comprehensible) than Korz.

While mdp rests on a Prolog, a single-dimensional language, its unique properties of uniformity, minimality and homoiconicity provide a flexible foundation for embedding multidimensionality and experimenting with new features. This allowed us to quickly implement one of Ungar et al.'s suggested improvements to Korz, selector-based matching (anonymous rules). Also, unlike Korz, multiple methods (predicates) can be executed if they have the same specificity. We made this design choice because non-determism is an important and desired feature of Prolog programming and it naturally extends to multidimensional predicates. Another notable difference is mdp's reliance on weights. The initial design did not include weights, but when introducing subtyping, it became apparent that there must be a kind of weight dimension.
Also, we can exploit Prolog's unification capabilities to perform pattern matching on dimension coordinates which will be explored in future works.

\section{Conclusion and Future Work}

We have seen that we can use mdp for many purposes. We can use multidimensional predicates to do object-oriented programming and to extend a program with additional concerns. Using techniques such as memoization, we obtain the benefit of having readable and efficient specifications. Yet we have barely scratched the surface and we will present more scenarios in future works. 

In particular, there are following issues that need attention:

\begin{itemize}

\item \textbf{Efficiency of Implementation} 

A major drawback of mdp is its efficiency. The dispatching mechanism for multidimensional predicates is very inefficient and has been programmed with quick adaptability and readability in mind. Its performance is unlikely to be satisfactory in production systems.
For satisfactory performance, a Prolog VM could be extended with special instructions for multidimensional dispatching. 

However, it is not clear at this stage whether the mechanism is sufficiently expressive for large-scale Prolog systems. To this end, more scenarios have to be investigated to reveal possible shortcomings. 

\item \textbf{Debugging and Comprehensibility} 

Currently, there is no support for debugging mdp programs. When debugging an mdp program, the programmer has to trace the call-chain of the mdp implementation to get information such as the context and predicate selection. A usable debugger has to hide these implementation details from the user and present the desired information in an easily accessible way. 

Also, the programmer has to be very careful in organizing multidimensional predicates. When programming in the large-scale, a development environment that helps a programmer organizing dimensions and point out at possible mistakes is mandatory.

The design of such a development environment with a suitable debugger will be an interesting challenge.

\item \textbf{Weight Dimension and Pattern Matching on Dimension Values}

We have seen that in many cases such as implementing subtyping, a weight dimension is necessary to ensure selection of the most specific subtype. However, instead of using annotations, a cleaner solution would be to explicitly define a weight dimension and use pattern matching to determine the weights of rules: 

\begin{lstlisting}
[weight: 1] # p.
[weight: 2] # p.

?- [weight: X] ? p
	X = 1 ;
	X = 2.
\end{lstlisting} 

An anonymous could then do weight-based predicate selection. This would further simplify the matching rules and the dispatcher. However, one problem is that in our example both variants are executed. In our case this is not desirable, but there might be cases in which we want to match on dimensions and on arguments of the predicate. Experiments have to provide an answer which approach is practical.

\item \textbf{Extending the mop}

The meta-object protocol we presented is far from complete. For example, we can enrich mop with syntactical sugar for defining OOP-style properties and context-oriented visibility modifiers. Also, the syntax for matching subtypes is a bit verbose, i.e., \verb|[rcvr: OID, OID < Subtype]| can be shortened to \verb|[OID < Subtype]| (like in Korz). Modifying mop accordingly is straightforward. 

\item \textbf{Restructuring Prolog Libraries}

mdp gives us many new ways to structure Prolog code and much incentive to restructure existing code bases. For example, the \verb|ordsets|-library provides special predicates for manipulating and querying data types in ordered sets such as \emph{ord\_intersection/3} or \emph{ord\_memberchk/2}. With mdp we can give the programmer a more consistent experience and have overloaded predicate names such as \emph{member} and, depending on the context, dynamically choose the predicates to deal with special data types.

Another interesting experiment would be to rewrite XPCE \cite{xpce}, a GUI framework for SWI-Prolog, using the techniques we have presented in this paper.

\item \textbf{Correctness}

Currently, mdp is an experimental stage, so bugs are to be expected even though we have made sure that the presented examples work. To make mdp ready for production systems, rigorous testing has to be applied along with formal verification, where we will take advantage of Prolog's well-defined language semantics.

\end{itemize}

We welcome suggestions, bug reports and improvements to the implementation which can be obtained at: \url{https://bitbucket.org/nexialist/mdp}.

\bibliographystyle{alpha}
\bibliography{references}

\begin{thebibliography}{UOK14}

\bibitem[CUS95]{object-focus}
Bay-Wei Chang, David Ungar, , and Randall~B. Smith.
\newblock {\em Getting Close to Objects: Object-Focused Programming
  Environments}.
\newblock In Visual Object-Oriented Programming, Margaret Burnett, Adele
  Goldberg, and Ted Lewis, eds., Prentice-Hall, 1995.

\bibitem[HO93]{subject-oriented-programming}
William Harrison and Harold Ossher.
\newblock Subject-oriented programming: A critique of pure objects.
\newblock {\em SIGPLAN Not.}, 28(10):411--428, October 1993.

\bibitem[LRN11]{subject-topia}
Daniel Langone, Jorge Ressia, and Oscar Nierstrasz.
\newblock Unifying subjectivity.
\newblock In {\em Objects, Models, Components, Patterns - 49th International
  Conference, {TOOLS} 2011, Zurich, Switzerland, June 28-30, 2011.
  Proceedings}, pages 115--130, 2011.

\bibitem[UOK14]{korz}
David Ungar, Harold Ossher, and Doug Kimelman.
\newblock Korz: Simple, symmetric, subjective, context-oriented programming.
\newblock In {\em Proceedings of the 2014 ACM International Symposium on New
  Ideas, New Paradigms, and Reflections on Programming \& Software}, Onward!
  2014, pages 113--131, New York, NY, USA, 2014. ACM.

\bibitem[US87]{self}
David Ungar and Randall~B. Smith.
\newblock Self: The power of simplicity.
\newblock In {\em Conference Proceedings on Object-oriented Programming
  Systems, Languages and Applications}, OOPSLA '87, pages 227--242, New York,
  NY, USA, 1987. ACM.

\bibitem[WA02]{xpce}
Jan Wielemaker and Anjo Anjewierden.
\newblock Programming in {XPCE/Prolog}, 2002.
\newblock Available online:
  http://info.ee.pw.edu.pl/Prolog/Download/userguide.pdf, Accessed: 07.03.2016.

\end{thebibliography}

\end{document}